\documentclass[useAMS]{mn2e}
\usepackage{graphicx}
\usepackage{epsfig}
\title{High Latitude Dust and the 617 MHz - 850 $\mu$m relation in NGC 5775}
\author[R.\ S.\ Brar et al.]
       {Rupinder Singh Brar,$^{1}$ Judith A.\ Irwin$^{1}$ and D.\ J.\ Saikia$^{2}$\\
        $^{1}$Queen's University, Physics Department, Kingston, ON, K7L 3N6 Canada \\
        $^{2}$National Centre for Radio Astrophysics, TIFR, Post Bag 3, Ganeshkhind, 
              Pune 411 007, India}
\date{}
     
\pagerange{\pageref{firstpage}--\pageref{lastpage}}
\pubyear{2002}

\begin{document}
    
\maketitle
    
\label{firstpage}
   
\begin{abstract}
We present continuum observations of the edge-on spiral
galaxy NGC 5775 at 617 MHz with the Giant Metrewave Radio Telescope (GMRT) 
and at 850 $\mu$m with the Submillimetre Common-User Bolometer Array 
(SCUBA) on the James Clerk Maxwell Telescope (JCMT).  We report the detection
of dust at high-latitudes, extending to about 5 kpc from the disk of the galaxy,
and spurs of non-thermal radio emission.
The 617 MHz and 850 $\mu$m  distributions are compared with one another and 
we find a strong correlation between metre wavelength radio emission and sub-mm 
emission in the disk and at high-latitudes. This suggests that there may be
a more fundamental relationship between cold dust and synchrotron radiation 
other than the link via star formation.
\end{abstract}

    
\section{Introduction}
    
There is now considerable evidence for galactic halos, thick disks,
and discrete connecting features between the disk and the halo in many
star-forming spiral galaxies.  Such features are best detected directly by
observing edge-on galaxies with sufficient sensitivity and spatial
resolution.  High-latitude (i.e. $\geq$ 1 kpc) emission has now been
observed in every interstellar medium (ISM) tracer (cf. Dahlem 1997).  
NGC 5775 is an example of a galaxy in which every ISM component has been 
observed at high-latitudes (Lee et al. 2001).

The Infrared Astronomical Satellite (IRAS) data of a sample of
spiral galaxies showed the estimated ratio of their gas mass to 
dust mass, to be about 10 times larger than the average value
for the inner region of the Galaxy (Devereux \& Young 1990).
It was suggested that this could be due to a large amount of cold dust
in these galaxies that IRAS was not sensitive to.  Observations
with good spectral coverage and sensitivity were required 
to disentangle the warm and cold dust components.  
The Submillimetre Common-User Bolometer Array (SCUBA) operating on the
James Clerk Maxwell Telescope (JCMT), which we have used for this project,
allows astronomers to determine the distribution
and quantity of such cold dust with high resolution.
Currently, only  a small number
of quiescent and active edge-on spiral galaxies have had their dust 
distributions mapped using either SCUBA or the IRAM 30-m telescope; these include 
NGC 891 (Alton et al. 1998), NGC 3079
(Stevens \& Gear 2000), NGC 4631 (Neininger \& Dumke 1999), 
NGC 4565 (Neininger et al. 1996), NGC 5907 (Dumke et al. 1997) 
M82, NGC 253, NGC 4631 (Alton et al. 1999) and now NGC 5775.

In the 1970s a correlation was discovered between mid-infrared 
($10\mu$m) and radio (1425 MHz)
luminosities from the nuclei of Seyfert galaxies, and later normal
spiral galaxies (van der Kruit 1971, 1973; Rieke 1978).  In the 1980s the IRAS 
mission established a universal and tight correlation between the far infrared 
(FIR) and the radio continuum luminosities of galaxies, 
(de Jong et al. 1985; Helou et al. 1985).  Recently a number of
authors have investigated the FIR/radio correlation within the disks 
of nearby galaxies in an attempt to gather more detailed information
on how the correlation works (Beck \& Golla 1988; Bicay \& Helou 1990, 
amongst others).

Presently, it is widely believed that the main reason behind 
the FIR/radio correlation, is that both emissions depend on the same recent
star formation activity of that particular galaxy (Condon 1992).
That is, hot young stars heat the dust, producing FIR emission,
as well as ionize gas and provide a supply of SNe, producing the radio continuum
emission.  However, the situation may not be so clear cut. For example, in
a detailed study of M31, Hoernes, Berkhuijsen \& Xu (1998, hereinafter referred
to as HBX), reported the discovery of a nonthermal-radio/cool-dust correlation, 
quite separate from the thermal-radio/warm-dust correlation
that is linked through massive ionizing stars. It is worth noting that 
HBX did not find evidence of a correlation between either
non-thermal radio with warm dust, or thermal radio with cool dust.
Thermal emission from warm dust is the result of heating from 
massive stars ($>20 M_{\odot}$), while cool dust could be powered by 
intermediate mass stars (5-20 $M_{\odot}$) and the interstellar radiation field.  
The nonthermal-radio/cool-dust 
correlation could not be explained through a dependence on the same 
energy source.  HBX suggest a possible scenario involving coupling of gas, 
cold dust and magnetic fields.

In this paper we present the first detection of high-latitude dust for NGC 
5775.  Observing with SCUBA, we map the cold dust in the disk and halo. We also present 
radio continuum observations of the galaxy
taken with the Giant Metrewave Radio 
Telescope (GMRT), which trace largely the non-thermal synchrotron emission.
The observations from SCUBA and GMRT are spatially compared with
one another in an effort to investigate whether a cold dust/non-thermal radio
emission correlation exists in the disk and halo of this galaxy.
NGC 5775 is an edge-on spiral galaxy at a distance of 24.8 Mpc 
(H$_o$=75 km s$^{-1}$ Mpc$^{-1}$), and  
with an inclination angle of 86$^\circ$ (Irwin 1994).  NGC 5775 is found on the 
FIR/radio correlation at log P$_{20 cm}$ = 22.32 W Hz$^{-1}$ and 
log (L$_{FIR}$/L$_\odot$) = 10.44.
The results in this paper are the first ones of an ongoing project
to investigate a possible sub-mm/metre wavelength correlation
in the disks and halos of galaxies.

\section[]{Observations and Data Reduction}

\subsection{Submillimetre}
Observations of NGC 5775 were made on 2001 January 4 and 5,
at the JCMT using SCUBA.  SCUBA images
a $2.3^{\prime}$ region of sky simultaneously at 450 and 850 $\mu$m.
Our observations
provide fully sampled images by moving the secondary mirror in the 64pt jiggle
map mode.  We performed pointing checks every hour against a bright source.
Skydip measurements were also made to determine the atmospheric transparency by 
change in elevation (opacity
was about 0.27 at 850 $\mu$m and 1.67 at 450 $\mu$m).  Mars was mapped twice
each night for calibration of our source and to determine the beam shape.  
Our observations of NGC 5775 consisted of three exposures of 45 minutes each over 
two overlapping fields, since our galaxy is larger than the diameter
of the region that the telescope can image.

The software package SURF was used to reduce the SCUBA data.  All maps were
reduced using typical reduction tasks.  SURF was used to clean and flat-field 
the images.  Dirty bolometers were removed and the image noise was reduced 
using a SURF task which compensated for spatially correlated emission. 
The pointing drifts were corrected and the images were calibrated using Mars.
Finally all the observations were combined into a single map.  The resulting
map has a resolution of 14.6$^{\prime\prime}$ and an rms noise of 4 mJy/beam.

\subsection{Radio Continuum}
NGC 5775 was observed with the GMRT on 2000 July 1 and 16 at 617 MHz
using a bandwidth of 16 MHz.  The GMRT consists of 
thirty steerable, 45-metre diameter antennas, of which 14 are located 
randomly in a central 1 km $\times$ 1 km diameter region referred to
as the Central Square.  The remaining antennas lie along 3 arms of an
approximately `Y' configuration, with each arm extending to a length of 
about 14 km.  This provides $uv$ coverage for imaging both the compact 
and extended emission.  A more detailed description of the GMRT can be
found in Swarup et al. (1991).  These observations were made during the
commissioning phase of the GMRT and typically 25 to 27 antennas were 
available for each day.  The source was observed on two different days for 
checking consistency of the images.  The primary flux density calibrator 
was 3C286 while the phase calibrator was 3C298.  The total time spent
on the source was approximately 120 minutes on each observing day.

The analysis of the data was done using the Astronomical Image 
Processing System (AIPS) from the National Radio Astronomy
Observatory.  The data editing itself was done within the AIPS 
environment using the task UVFLG. However, the identification 
of bad data was done outside the AIPS environment with 
algorithms developed by J. Stil in a package called `borg'.
The borg allows the user to define what constitutes bad data.  
For our data, borg identified all points that had near zero
flux density at each channel for the flux density and phase calibrator.
The routine also identified the spikes in the data while observing the
calibrators by finding points that deviate more than, approximately 3 sigma 
from the modal amplitude for a particular baseline, channel and source.  
If any baseline had a significant number of points flagged, it was
subsequently flagged from all sources. The borg package helped to ease
the editing of the data because the GMRT did not have any automatic flagging
of bad data, including bad antennas at the time of these observations. 

The AIPS tasks CALIB, CLCAL and BPASS were used on the
flagged data set to calibrate the source.  The central 100 
channels of the total of 128 channels were used in the analysis.
The channels were averaged and  Fourier inversion and cleaning were
performed on the NGC 5775 data using IMAGR. A series of self-calibration
and cleaning procedures were then conducted to produce two maps, one from
each day of observations.  After determining that the two maps were virtually
identical we used the AIPS task DBCON to combine the UV plane data from each 
observation and produced a final single map to increase our signal-to-noise ratio.  
Using the brightest source visible in our observation and the rms noise
of the map, 0.9 mJy/beam, we calculate a dynamic range of 112.  Although 
the angular resolution using the entire GMRT is approximately 6 arcsec, 
we have tapered the GMRT images to 14.6$^{\prime\prime}$ 
to be comparable with the 850 $\mu$m SCUBA map.  

\section{Results}

\subsection {Submillimetre and radio maps}
In Fig. 1 we show our final 850 $\mu$m map of the edge-on spiral galaxy NGC 
5775.  The 450 $\mu$m map (not shown) has a very similar disk morphology with 
poorer signal-to-noise ratio.  SCUBA maps of NGC 5775 reveal a bright central 
source and a number of additional peaks of bright emission, approximately along the disk
(Note that the figure-8 outline is an artifact due to combining the data 
from the two fields used.).

\begin{figure}
\centering \epsfig{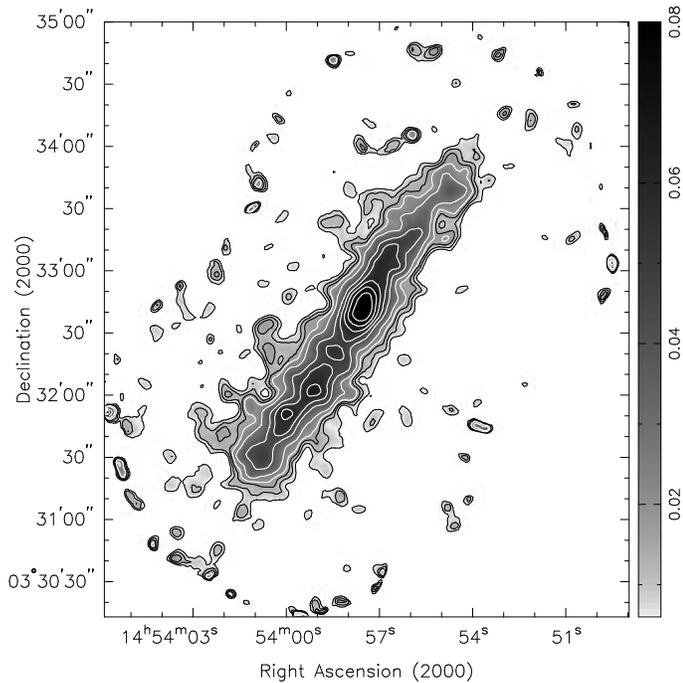}
 \caption{SCUBA map of NGC 5775.
 Grayscale levels are given at the right in Jy/beam.  The resolution is 14.6$^{\prime\prime}$.
The rms noise is 4 mJy/beam and the peak brightness is at 89 mJy/beam.  For clarity of features 
contours are shown at levels 6, 10, 14, 18, 26,
40, 50, 60, 70 and 80 mJy/beam.}
\end{figure}

One of the most striking features of the map is the
emission that rises from the disk of the galaxy into the halo.  
Clear evidence of emission is found at large
distances from the disk; in three cases at latitudes of about 5 kpc.  
This is among the farthest from the plane that dust has been 
detected for any galaxy.  Although the signal-to-noise ratio is modest, these
features are visible in separate observations on different days, providing supporting 
evidence of their reality (Brar \& Irwin 2002).

The 617 MHz radio-continuum map of NGC 5775 is presented in Fig. 2.  A
comparison of the GMRT image with a VLA 1400-MHz image with a similar angular
resolution (Lee et al. 2001) shows the emission in the disk to be very similar, 
but the spurs of emission extending from the disk are more clearly
discernible in the 617 MHz image.  The disk of NGC 5775
in radio continuum emission is spatially very similar to the SCUBA image shown in 
Fig. 1.  In particular, it possesses a central bright spot and a 
number of other bright areas of emission along the disk, including a
secondary peak towards the lower half of the galaxy.

Radio emission too is detected at high-latitudes.  The emission at 617 MHz, which
is predominantly non-thermal, is observed above the disk
along the entire galaxy, with the high-latitude features being particularly prominent 
on the northern side of the galaxy axis.  

\begin{figure}
\centering\epsfig{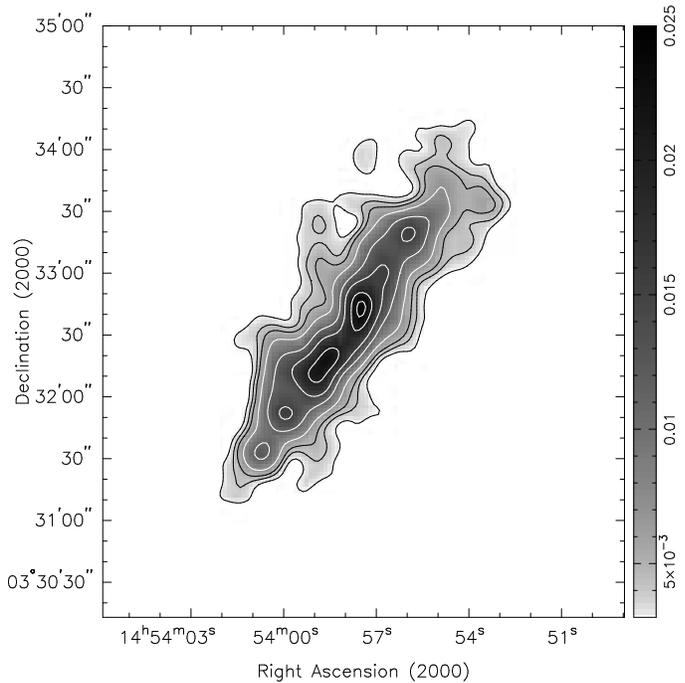}  
 \caption{GMRT map of NGC 5775.  The grayscale
 levels are given in Jy/beam.  The resolution is 14.6$^{\prime\prime}$.
The rms noise is 0.9 mJy/beam and the peak brightness is at 24 mJy/beam.  
For clarity of features contours are shown at levels 3, 4, 5, 7, 11, 15, 20 and 23 mJy/beam.}
\end{figure}

\subsection{Comparison of maps}

Fig. 3 is an optical image from the Digitized Sky Survey (DSS) 
with the 850 $\mu$m contour plot overlayed on it.  This overlay demonstrates the large 
distances that dust is detected away from the optical disk.  Within
the disk, the brightest regions of sub-mm emission do not correlate with  
the brightest regions of optical emission.  In fact, even the central bright
regions are slightly separated.  In the southern half of the galaxy, the regions
of high optical and dust emission appear to be anti-correlated, as expected
if the dust is obscuring starlight.

\begin{figure}
\centering\epsfig{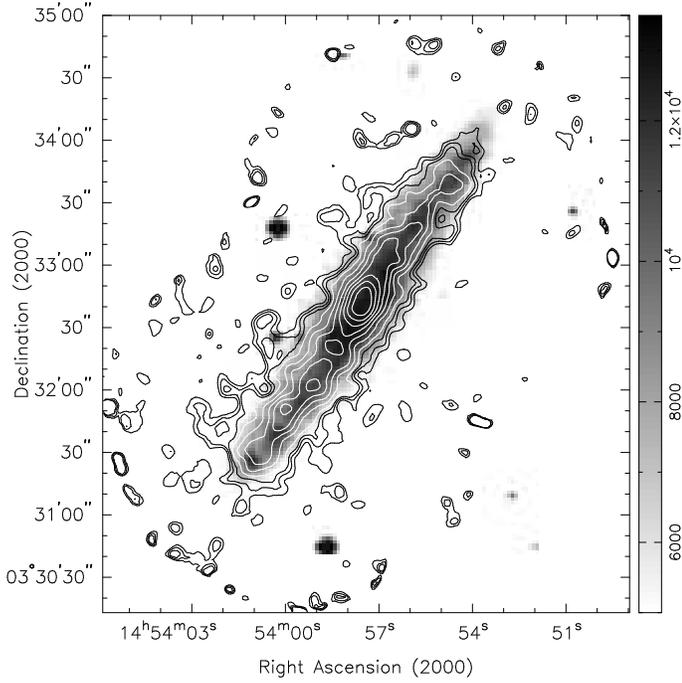}  
 \caption{The JCMT 850 $\mu$m contour map overlayed
 on a DSS optical image.  The contour levels are 6, 10, 14, 20, 30, 40, 50, 60, 70 
and 80 mJy/beam.}
\end{figure}

Fig. 4 shows the GMRT map overlayed on the SCUBA map.  The spatial
correlation of the features in the disk is high.  More surprisingly, we 
also find a similar correlation at high-latitudes.
Note that the two lower streams (14$^h$ 54$^m$ 02$^s$, 3$^\circ$ $32^{\prime}$) in 
the SCUBA map do have corresponding emission in the GMRT map, but are at lower 
brightness levels than demonstrated in Fig. 4 and are displayed separately in Fig. 5.  
In Fig. 5 there is a single black contour level at 1.4 mJy/beam.  Thus the contour 
located between the two features indicates a depression in radio emission.

\begin{figure}
\centering\epsfig{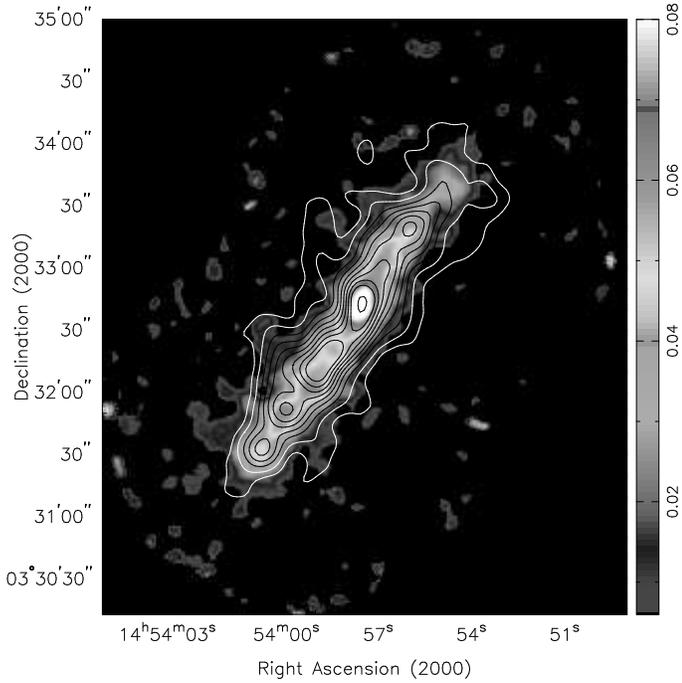}  
 \caption {The 617 MHz continuum contours overlayed on the 850 $\mu$m
 halftone map.  The contour levels are 3.3, 5, 7, 9, 11, 13, 
15, 17, 19 and 23 mJy/beam.  The 850 $\mu$m is presented as a cyclic
halftone in order to emphasize the features in the disk.}
\end{figure}

\begin{figure}
\centering\epsfig{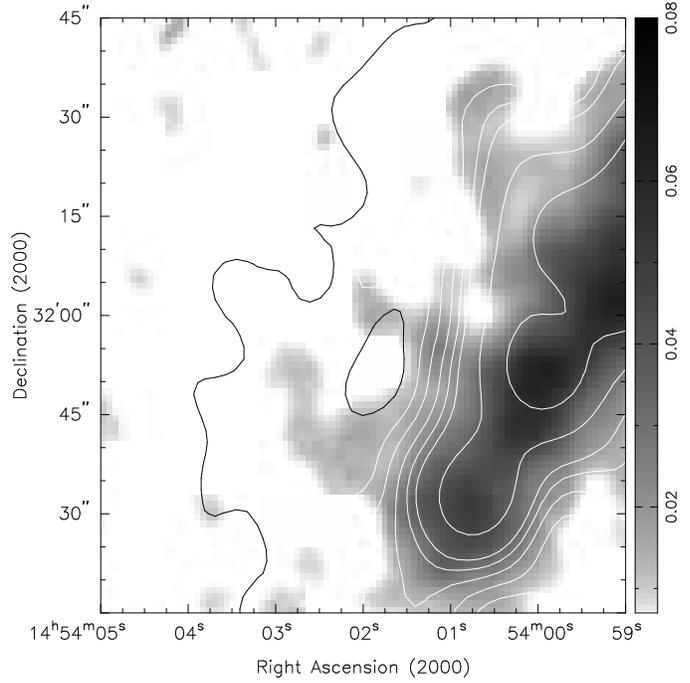}
 \caption {The 617 MHz continuum contours overlayed on the 850 $\mu$m
 grayscale map.  The white contour levels are 2.2, 3, 4, 5, 6, 9 and 13 mJy/beam.  
The lone black contour is 1.4 mJy/beam.  Notice the depression in radio emission 
between the two high-latitude features.}
\end{figure}

In Fig. 6 we present z profiles of the JCMT and GMRT maps, both above and below 
the plane.  A fairly typical profile was chosen to demonstrate the separation
of the halo and disk, the exact positions of which are given in Fig. 6.
Fig. 6 a) and b) profiles indicate evidence for a halo at
about 13 arcsec above (north-east) the maximum intensity in the disk.  
Fig. 6 c) and d) profiles show 
halo evidence beginning at about 11 arcsec below (south-west) the plane.  
This translates
into, evidence for a halo starting at between 1.32 and 1.56 kpc from the plane of NGC 5775.
From these profiles and a galaxy inclination of 86 degrees, we can safely state that the
features we are observing reside in the halo and not the disk.

\begin{figure}
\begin{center}
\begin{tabular}{cc}
 \epsfig{figure=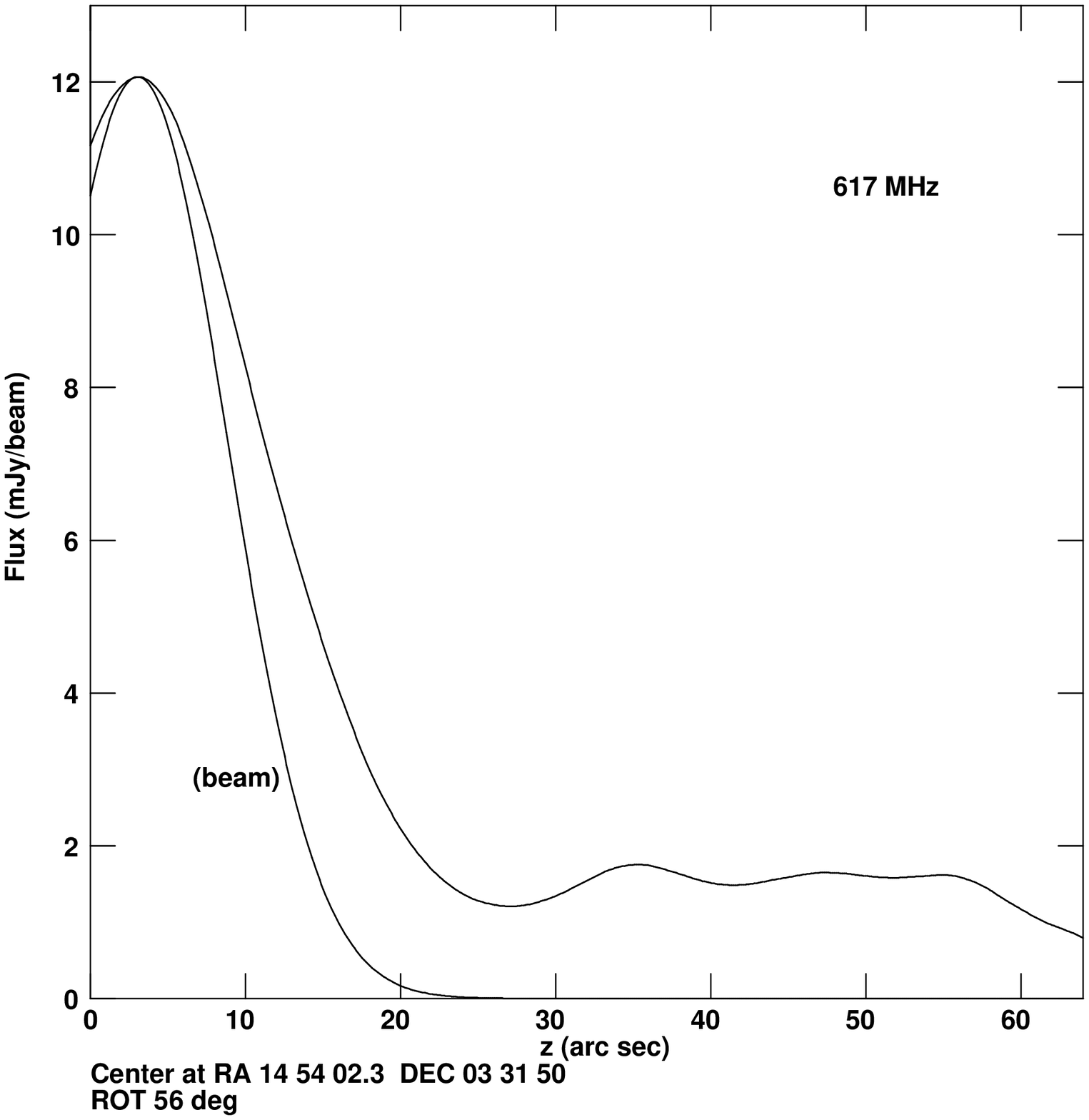, width=4.1cm}  a)&
 \epsfig{figure=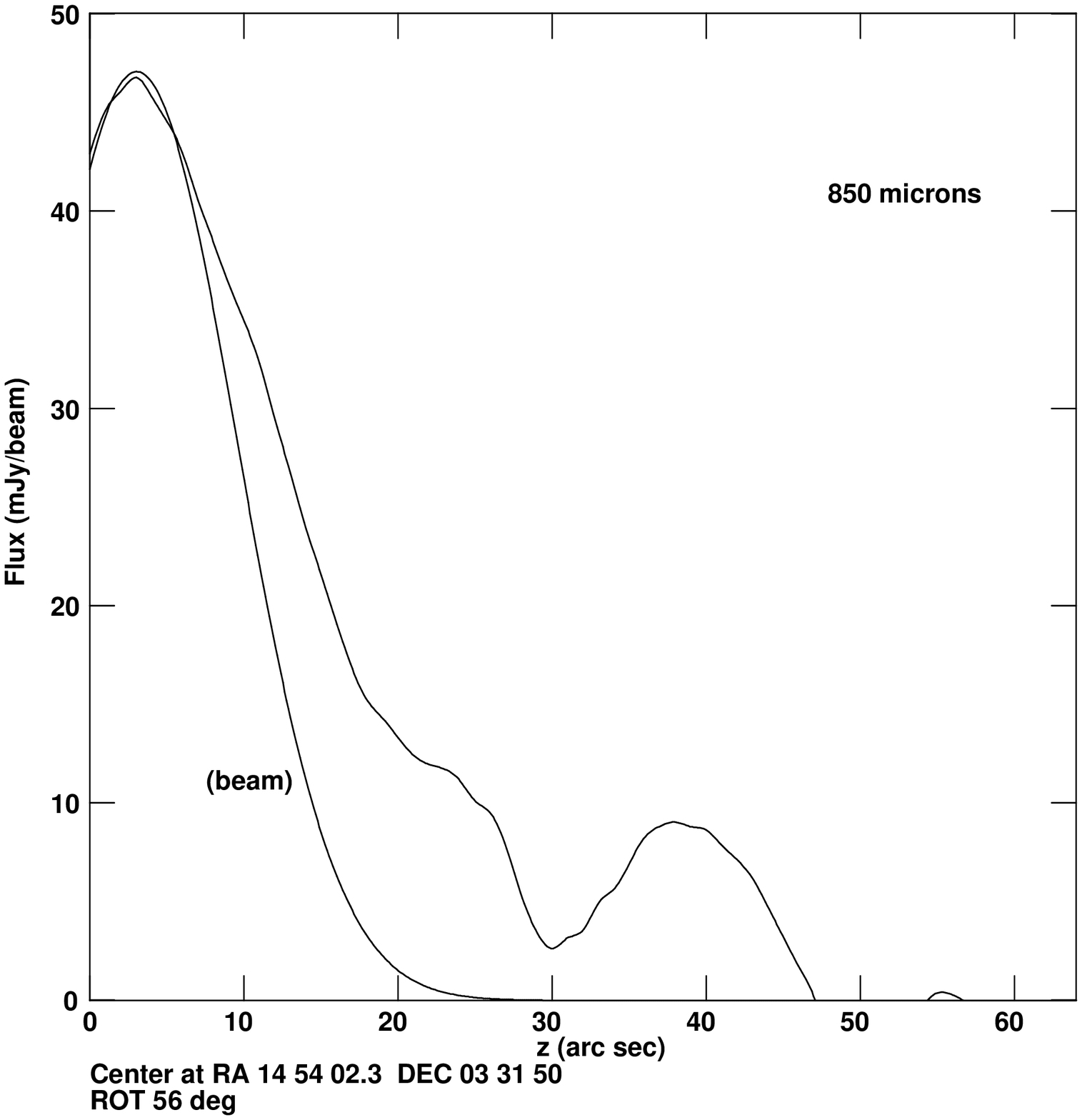, width=4.1cm}  b)\\
 \epsfig{figure=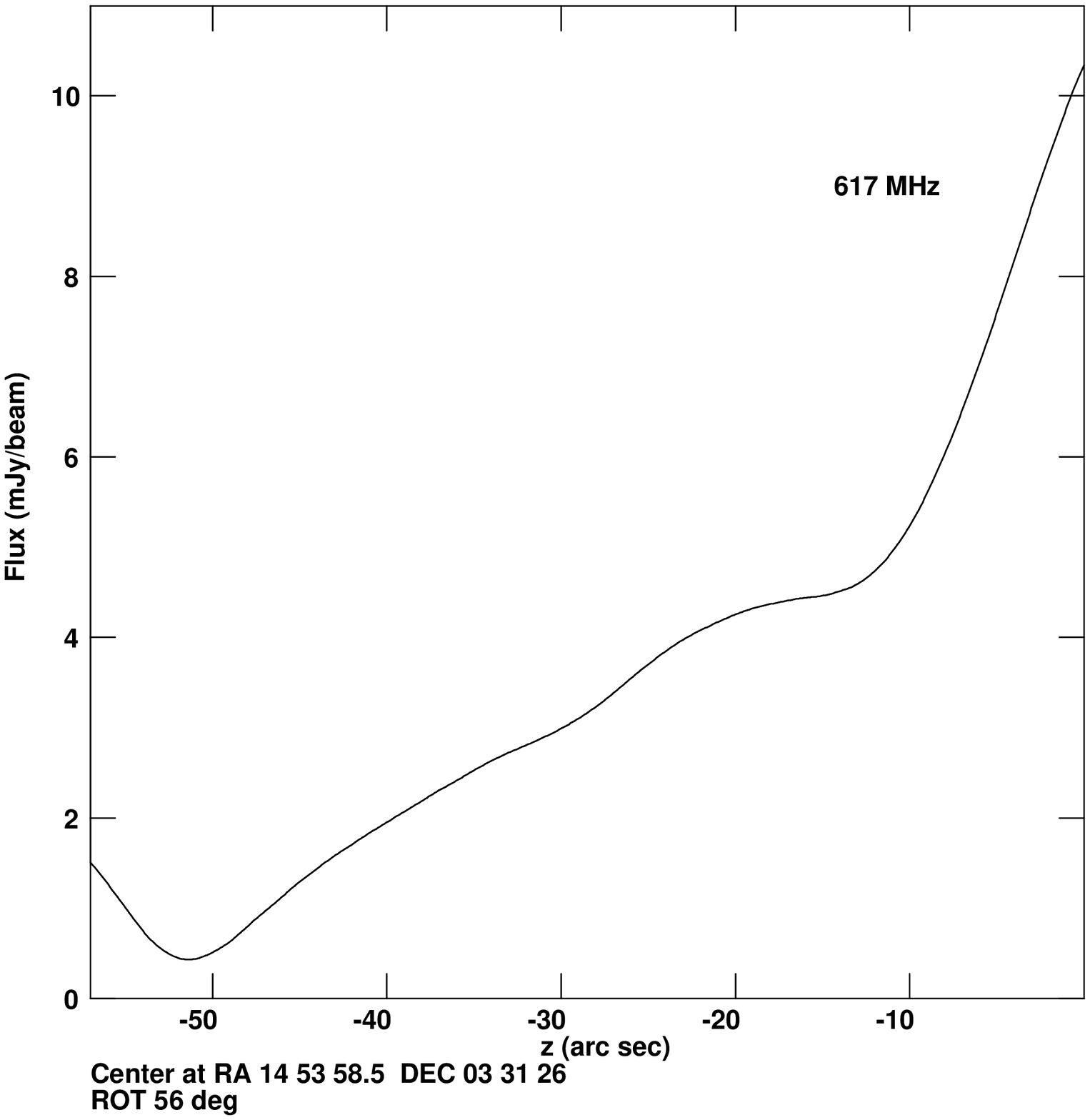, width=4.1cm}  c)&
 \epsfig{figure=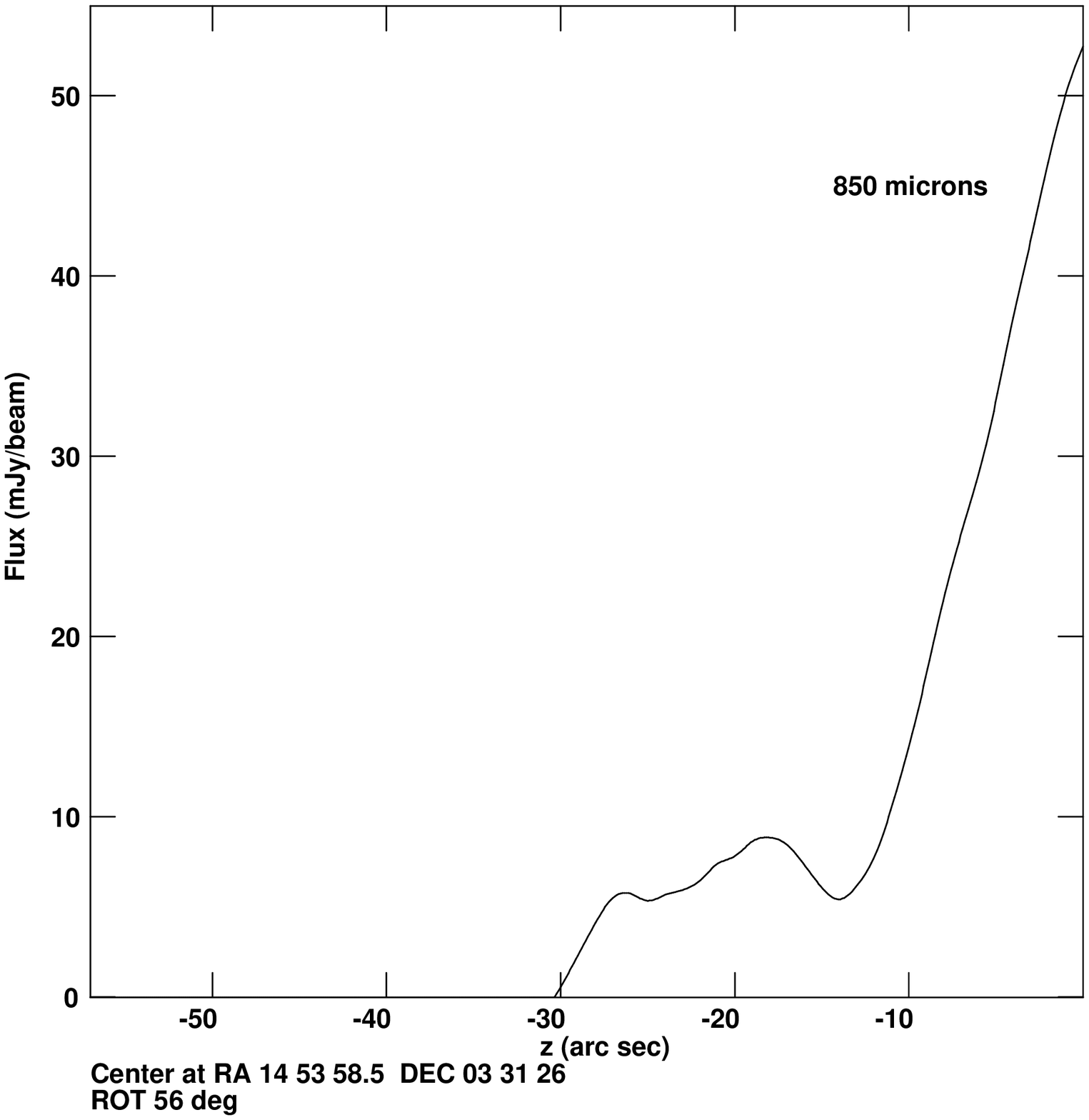, width=4.1cm}  d)\\
\end{tabular}
\end{center}
 \caption{The z profiles presented show flux in 
mJy/beam against distance from plane in arcsec.  The mid-plane positions are marked beneath each plot.
In a) and b) the beam shape of each observation is also plotted.  
a) 617 MHz profile above plane. b) 850 $\mu$m profile above plane, at identical location 
to a).  c) 617 MHz profile below plane. d) 850 $\mu$m profile below plane, 
at identical location to c).}
\end{figure}

\section{Discussion}
\subsection{Correlation}
The similarities between the sub-mm and radio maps demonstrate that
a correlation exists between the radio continuum and sub-mm radiation in NGC 5775.
A more quantitative demonstration is shown in Fig. 7, where 
we have plotted the flux densities at 850 $\mu$m against those at 617 MHz 
at identical independent positions for the entire disk and halo of NGC 5775.  
Fig. 8 shows the correlation for positions solely in the disk.  It is
clear that sub-mm and radio emission are highly correlated.
The slope of the linear-least squares fit to the entire disk and halo is
0.26$\pm$0.01 (Fig. 7), while for the disk alone the slope is $0.25\pm0.03$ (Fig. 8).  
The corresponding values for the north-eastern and south-western halos are $0.31\pm0.04$  
and $0.26\pm0.03$ respectively (Fig. 9). Within the errors, the slopes for   
each section of the galaxy are similar, suggesting that the reason behind this
correlation for different parts of the galaxy must also be similar.  This result
also suggests that the discrete kpc disk halo features originate in the disk of the
galaxy.  The fainter dotted line in Fig. 8 is the best fit of a power law regression, 
which is similar to the linear regression.  The slope of the power law is 
$0.85 \pm 0.08$.  This value is very close to one, indicating
the relationship is approximately linear.  For comparison, the slopes
typically found for the radio-FIR correlation are between 1.1 and 1.3.  

The sub-mm observations map the cold dust, which
emits predominantly longward of a couple of hundred $\mu$m (Sauvage 1997).
The detection of cold dust was not a surprising finding since dust-to-gas 
mass ratio of external galaxies derived from IRAS were approximately 10 times 
lower than in our galaxy (cf. Devereux \& Young 1990).  The IRAS observations
are largely sensitive to warm dust and helped establish the tight and universal
correlation between FIR and radio continuum emission. While this correlation
is attributed to the dependence of both emission processes on recent star
formation activity, the 617 MHz/850 $\mu$m correlation demonstrated here is unlikely
to be the result of a common energy source for the emission at the two wavebands.
The cold dust in galaxies is heated not by massive ionizing or intermediate mass 
non-ionizing stars, but possibly by the general interstellar radiation field 
(Hippelein et al. 2000).  Also, the radio map at 617 MHz, is almost entirely due 
to non-thermal synchrotron radiation.  From Condon (1992) we estimate the 
ratio of thermal flux density to total flux density at 617 MHz in NGC 5775 to be 
approximately 0.06 using a spectral index of 0.93 (Irwin et al. 1999).  The ratio of 
the thermal component is approximately twice the above value at 1420 MHz.

\begin{figure}
\centering\epsfig{figure=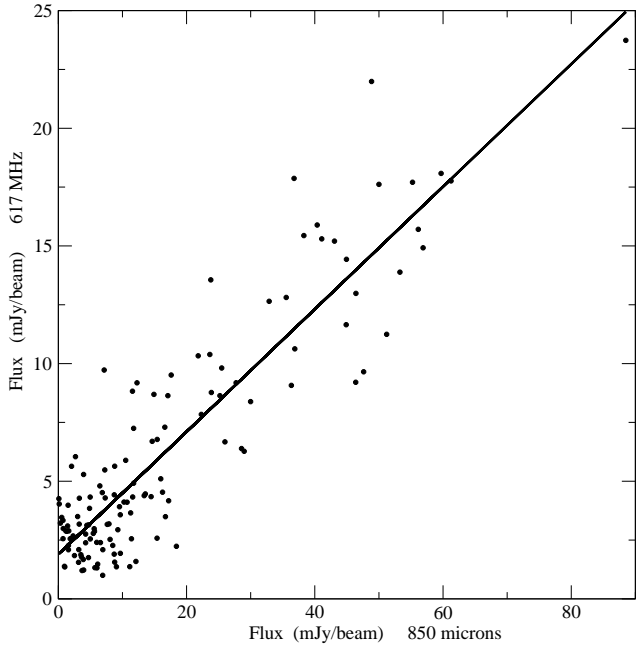, height=9cm}  
 \caption {The 850 $\mu$m flux density plotted against 617 MHz flux density for the
disk and halo of NGC 5775. The line indicates the linear least-squares fit to the data.}
\end{figure}

Further supporting evidence
that the correlation is not due to a simple shared energy source, comes from
the halo region of both maps.  The overlay in Fig. 4 demonstrates
that there is a spatial relation between the sub-mm and metre-wave radio emission
in the halo region of the galaxy.  This relation is more clearly shown in Fig. 9,
which is similar to Fig. 7 except that only high-latitude emission ($>$ 1.5 kpc) is 
shown.  The circles represent positions located in the halo on the north-eastern side of 
the galaxy axis, while the plus signs mark positions in the south-western halo.
In these regions there is no evidence of significant star formation. The link between the
two types of emission must be due to another factor.  

\begin{figure}
\centering\epsfig{figure=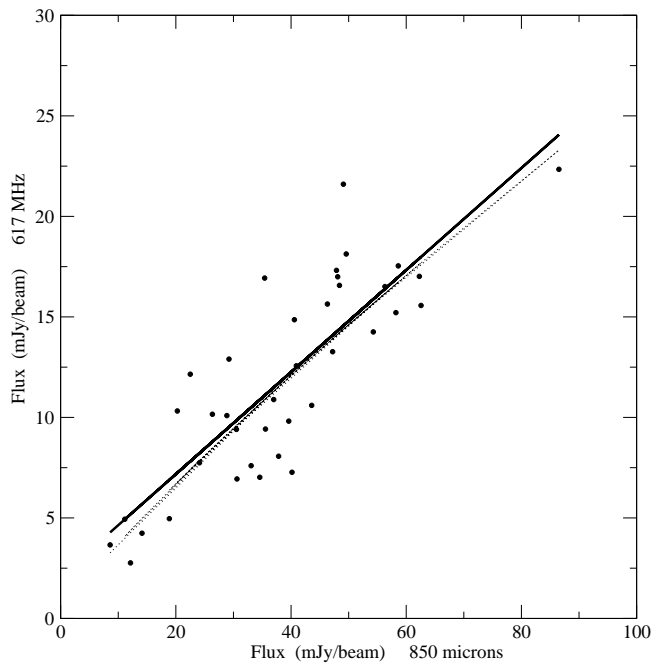, height=9cm}
 \caption{The 850 $\mu$m flux density plotted against 617 MHz flux density for the
disk of NGC 5775. The solid line indicates the linear least-squares fit to the data. The 
dotted line shows the best fit for a power-law relationship.}
\end{figure}

\begin{figure}
\centering\epsfig{figure=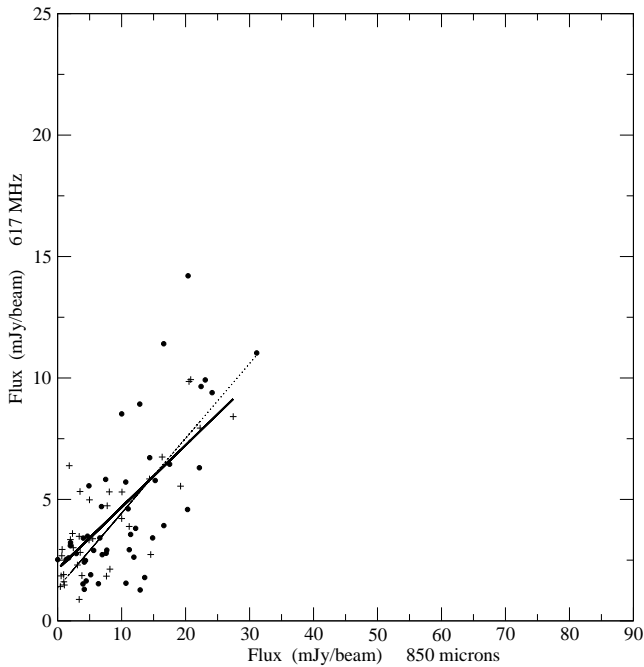, height=9cm}  
 \caption{The 850 $\mu$m flux density plotted against 617 MHz flux density for the
halo region. The halo on the north-east side of the galaxy axis of NGC 5775 are 
represented by circles, and the linear least-squares fit by the dotted line.  
The halo on the south-west side of the galaxy axis is
represented by the plus sign, and the least-squares fit by the solid line.}
\end{figure}

Our findings lead us to suggest that the well-known FIR/radio correlation
might be one aspect of a more fundamental relationship.  These results on NGC 5775 not only
demonstrate a relationship between the two types of emission, namely thermal emission
from cold dust and non-thermal emission from relativistic electrons, but could provide clues
towards understanding the distributions and relationships between gas, dust and magnetic
fields.  Studying the more abundant cold dust component, which could 
contribute approximately 80 to 95\% of the dust in normal spiral galaxies (Sievers et al. 
1994), has revealed a correlation between the non-thermal radio and sub-mm 
emission in NGC 5775. This could indicate a wider and more important relationship than one
based on recent star-formation activity alone.

Currently a model that can perhaps explain the correlation 
presented in this paper is the magnetic field-gas coupling model 
described by HBX to explain the relationship between the non-thermal and cold-dust
emission in M31.  In this model the cold dust and the non-thermal radio continuum emission 
are linked through the magnetic field coupled to the neutral gas which is mixed with 
the cold dust.  Using this model, HBX could explain the slope of $0.80 \pm 0.09$
for the observed relationship between the non-thermal radio and cool dust emission in M31 
(HBX). This slope is very close to our value of $0.85 \pm 0.08$ for NGC 5775.

An investigation of the magnetic field in NGC 5775 was conducted by
T\"{u}llmann et al. (2000).  An analysis of polarized radio continuum emission
at 1.49 and 4.86 GHz indicated that the magnetic field in the halo has
a strong component that is aligned with radio continuum spurs and 
are in a direction perpendicular to the disk of NGC 5775.  The magnetic fields
extend to sufficient latitudes from the disk and could provide an explanation
of the extra-planar features presented above at 850 $\mu$m and 617 MHz.
This supports the possibility that a magnetic field-gas coupling model
may explain the cold dust non-thermal gas correlation we have found in NGC 5775.

It is interesting to note that NGC 5775 may be interacting with the nearby 
galaxy NGC 5774 (Lee et al. 2001), and enquire whether the dust which has been 
detected up to $\approx$5 kpc feeds the intergalactic medium. However, without an 
estimate of the the velocity of the dust we cannot be certain that these dust grains 
will escape NGC 5775's gravitational well.  However, a few of the high-latitude 
features appear to suggest that dust might be falling back towards the disk.  
This is interesting in light of observations by Lisenfeld et al. (2002) which 
show no dust in the halo of the dwarf galaxy NGC 1569.  The mass difference between
dwarf galaxies and normal spirals may account for the missing dust, if the dust escaped 
into the intergalactic medium.

\section{Summary}
We have presented maps of NGC 5775 at 850 $\mu$m which are due to emission from cold dust
and at 617 MHz where the emission is due to synchrotron emission from relativistic electrons.
We detect dust at high-latitudes extending up to approximately 5 kpc, and spurs of
non-thermal radio emission.  The discrete high-latitude features originate in the disk of 
the galaxy.  The non-thermal and sub-mm emission are correlated in the 
disk and as well as at high-latitudes.  This correlation is unlikely to be due to the
recent star formation activity in the galaxy, but possibly reflects a coupling of gas,
dust and magnetic fields in the galaxy.
This galaxy is the first one of an ongoing study that we are conducting
on high-latitude emission from edge-on spiral galaxies.
Other galaxies must be analyzed to determine whether the correlation
observed in NGC 5775 is more widespread.

\section*{Acknowledgments}

We would like to thank Jeroen Stil for creating the borg program, which
assisted us in flagging GMRT data.
We thank the staff of the GMRT that made these observations possible. GMRT
is run by the National Centre for Radio Astrophysics of the Tata Institute
of Fundamental Research.
The JCMT is operated by the Joint Astronomy
Centre on behalf of PPARC for the United Kingdom, the
Netherlands Organisation for Scientific Research, and the National
Research Council of Canada.

\bsp
 
\label{lastpage}

\end{document}